\documentclass[conference]{IEEEtran}
\IEEEoverridecommandlockouts
\usepackage{ragged2e}
\usepackage{cite}
\usepackage{tikz}
\usetikzlibrary{positioning}
\usepackage{amsmath,amssymb,amsfonts}
\usepackage{algorithmic}
\usepackage{graphicx}
\usepackage[colorlinks,citecolor=red,urlcolor=blue,bookmarks=false,hypertexnames=true]{hyperref} 
\usepackage{textcomp}
\usepackage{xcolor}
\def\BibTeX{{\rm B\kern-.05em{\sc i\kern-.025em b}\kern-.08em
    T\kern-.1667em\lower.7ex\hbox{E}\kern-.125emX}}
\begin{document}

\title{Drug Recommendation System based on Sentiment Analysis of Drug Reviews using Machine Learning \\
}

\author{\IEEEauthorblockN{\textsuperscript{} Satvik Garg}
\IEEEauthorblockA{\textit{Department of Computer Science} \\
\textit{Jaypee University of Information Technology}\\
Solan, India \\
satvikgarg27@gmail.com}
}
\maketitle

\begin{abstract}
Since coronavirus has shown up, inaccessibility of legitimate clinical resources is at its peak, like the shortage of specialists and healthcare workers, lack of proper equipment and medicines etc. The entire medical fraternity is in distress, which results in numerous individual's demise. Due to unavailability, individuals started taking medication independently without appropriate consultation, making the health condition worse than usual. As of late, machine learning has been valuable in numerous applications, and there is an increase in innovative work for automation. This paper intends to present a drug recommender system that can drastically reduce specialists heap. In this research, we build a medicine recommendation system that uses patient reviews to predict the sentiment using various vectorization processes like Bow, TF-IDF, Word2Vec, and Manual Feature Analysis, which can help recommend the top drug for a given disease by different classification algorithms. The predicted sentiments were evaluated by precision, recall, f1score, accuracy, and AUC score. The results show that classifier LinearSVC using TF-IDF vectorization outperforms all other models with 93\% accuracy.
\end{abstract}
\section{Introduction}
With the number of coronavirus cases growing exponentially, the nations are facing a shortage of doctors, particularly in rural areas where the quantity of specialists is less compared to urban areas. A doctor takes roughly 6 to 12 years to procure the necessary qualifications. Thus, the number of doctors can't be expanded quickly in a short time frame. A Telemedicine framework ought to be energized as far as possible in this difficult time [1].

Clinical blunders are very regular nowadays. Over 200 thousand individuals in China and 100 thousand in the USA are affected every year because of prescription mistakes. Over 40\% medicine, specialists make mistakes while prescribing since specialists compose the solution as referenced by their knowledge, which is very restricted [2][3]. Choosing the top-level medication is significant for patients who need specialists that know wide-based information about microscopic organisms, antibacterial medications, and patients [6]. Every day a new study comes up with accompanying more drugs, tests, accessible for clinical staff every day. Accordingly, it turns out to be progressively challenging for doctors to choose which treatment or medications to give to a patient based on indications, past clinical history.

With the exponential development of the web and the web-based business industry, item reviews have become an imperative and integral factor for acquiring items worldwide. Individuals worldwide become adjusted to analyze reviews and websites first before settling on a choice to buy a thing. While most of past exploration zeroed in on rating expectation and proposals on the E-Commerce field, the territory of medical care or clinical therapies has been infrequently taken care of. There has been an expansion in the number of individuals worried about their well-being and finding a diagnosis online. As demonstrated in a Pew American Research center survey directed in 2013 [5], roughly 60\% of grown-ups searched online for health-related subjects, and around 35\% of users looked for diagnosing health conditions on the web. A medication recommender framework is truly vital with the goal that it can assist specialists and help patients to build their knowledge of drugs on specific health conditions.

A recommender framework is a customary system that proposes an item to the user, dependent on their advantage and necessity. These frameworks employ the customers' surveys to break down their sentiment and suggest a recommendation for their exact need. In the drug recommender system, medicine is offered on a specific condition dependent on patient reviews using sentiment analysis and feature engineering. Sentiment analysis is a progression of strategies, methods, and tools for distinguishing and extracting emotional data, such as opinion and attitudes, from language [7]. On the other hand, Featuring engineering is the process of making more features from the existing ones; it improves the performance of models.

This examination work separated into five segments: Introduction area which provides a short insight concerning the need of this research, Related works segment gives a concise insight regarding the previous examinations on this area of study, Methodology part includes the methods adopted in this research, 
The Result segment evaluates applied model results using various metrics, the Discussion section contains limitations of the framework, and lastly, the conclusion section.
\section{Literature survey}
With a sharp increment in AI advancement, there has been an exertion in applying machine learning and deep learning strategies to recommender frameworks. These days, recommender frameworks are very regular in the travel industry, e-commerce, restaurant, and so forth. Unfortunately, there are a limited number of studies available in the field of drug proposal framework utilizing sentiment analysis on the grounds that the medication reviews are substantially more intricate to analyze as it incorporates clinical wordings like infection names, reactions, a synthetic names that used in the production of the drug [8].

The study [9] presents GalenOWL, a semantic-empowered online framework, to help specialists discover details on the medications. The paper depicts a framework that suggests drugs for a patient based on the patient's infection, sensitivities, and drug interactions. For empowering GalenOWL, clinical data and terminology first converted to ontological terms utilizing worldwide standards, such as ICD-10 and UNII, and then correctly combined with the clinical information.

Leilei Sun [10] examined large scale treatment records to locate the best treatment prescription for patients. The idea was to use an efficient semantic clustering algorithm estimating the similarities between treatment records. Likewise, the author created a framework to assess the adequacy of the suggested treatment. This structure can prescribe the best treatment regimens to new patients as per their demographic locations and medical complications. An Electronic Medical Record (EMR) of patients gathered from numerous clinics for testing. The result shows that this framework improves the cure rate.

In this research [11], multilingual sentiment analysis was performed using Naive Bayes and Recurrent Neural Network (RNN). Google translator API was used to convert multilingual tweets into the English language. The results exhibit that RNN with 95.34\% outperformed Naive Bayes, 77.21\%.

The study [12] is based on the fact that the recommended drug should depend upon the patient's capacity. For example, if the patient's immunity is low, at that point, reliable medicines ought to be recommended. Proposed a risk level classification method to identify the patient's immunity.  For example, in excess of 60 risk factors, hypertension, liquor addiction, and so forth have been adopted, which decide the patient's capacity to shield himself from infection. A web-based prototype system was also created, which uses a decision support system that helps doctors select first-line drugs.

Xiaohong Jiang et al. [13] examined three distinct algorithms, decision tree algorithm, support vector machine (SVM), and backpropagation neural network on treatment data. SVM was picked for the medication proposal module as it performed truly well in each of the three unique boundaries - model exactness, model proficiency, model versatility. Additionally, proposed the mistake check system to ensure analysis, precision and administration quality.

Mohammad Mehedi Hassan et al. [14] developed a cloud-assisted drug proposal (CADRE). As per patients' side effects, CADRE can suggest drugs with top-N related prescriptions. This proposed framework was initially founded on collaborative filtering techniques in which the medications are initially bunched into clusters as indicated by the functional description data. However, after considering its weaknesses like computationally costly, cold start, and information sparsity, the model is shifted to a cloud-helped approach using tensor decomposition for advancing the quality of experience of medication suggestion.

Considering the significance of hashtags in sentiment analysis, Jiugang Li et al. [15] constructed a hashtag recommender framework that utilizes the skip-gram model and applied convolutional neural networks (CNN) to learn semantic sentence vectors. These vectors use the features to classify hashtags using LSTM RNN. Results depict that this model beats the conventional models like SVM, Standard RNN. This exploration depends on the fact that it was undergoing regular AI methods like SVM and collaborative filtering techniques; the semantic features get lost, which has a vital influence in getting a decent expectation.

\section{Methodologies}
The dataset used in this research is Drug Review Dataset (Drugs.com) taken from the UCI ML repository [4]. This dataset contains six attributes, name of drug used (text), review (text) of a patient, condition (text) of a patient, useful count (numerical) which suggest the number of individuals who found the review helpful, date (date) of review entry, and a 10-star patient rating (numerical) determining overall patient contentment. It contains a total of 215063 instances.

Fig. 1 shows the proposed model used to build a medicine recommender system. It contains four stages, specifically, Data preparation, classification, evaluation, and Recommendation.

\begin{figure}[htbp]
\centerline{\includegraphics[width=7cm, height=8cm]{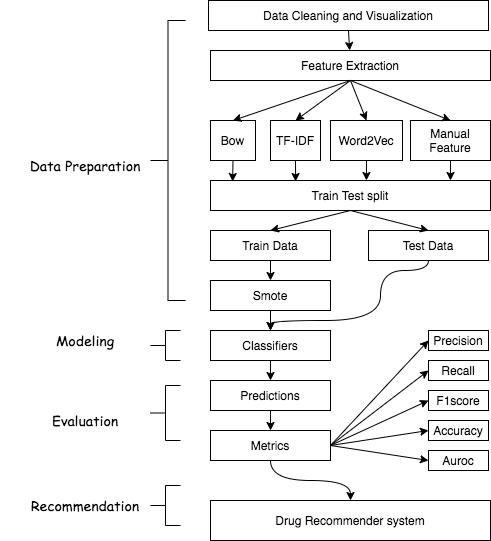}}
\caption{Flowchart of the proposed model}
\end{figure}
\subsection{Data Cleaning and Visualisation}
Applied standard Data preparation techniques like checking null values, duplicate rows, removing unnecessary values, and text from rows in this research. Subsequently, removed all 1200 null values rows in the conditions column, as shown in Fig. 2. We make sure that a unique id should be unique to remove duplicacy.
\begin{figure}[htbp]
\centerline{\includegraphics[width=8cm, height=4cm]{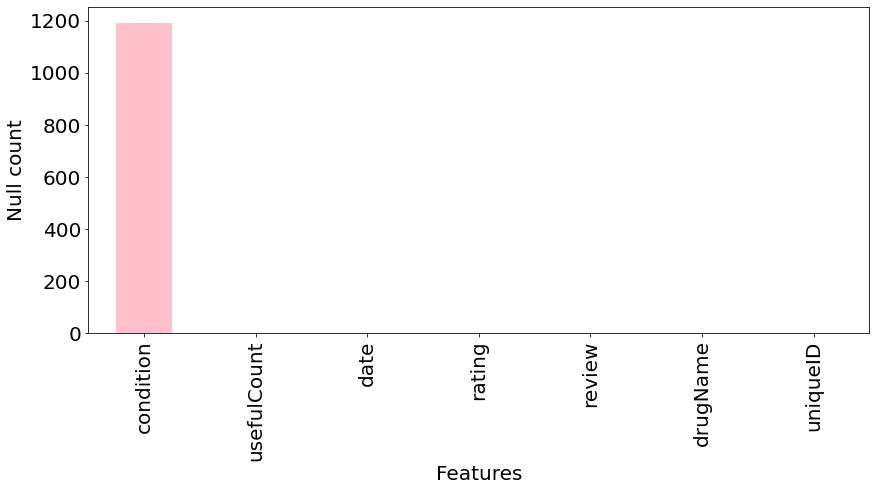}}
\caption{Bar plot of the number of null values versus attributes}

\end{figure}

Fig. 3 shows the top 20 conditions that have a maximum number of drugs available. One thing to notice in this figure is that there are two green-colored columns, which shows the conditions that have no meaning. The removal of all these sorts of conditions from final dataset makes the total row count equals to 212141.
\begin{figure}[htbp]
\centerline{\includegraphics[width=8cm, height=6cm]{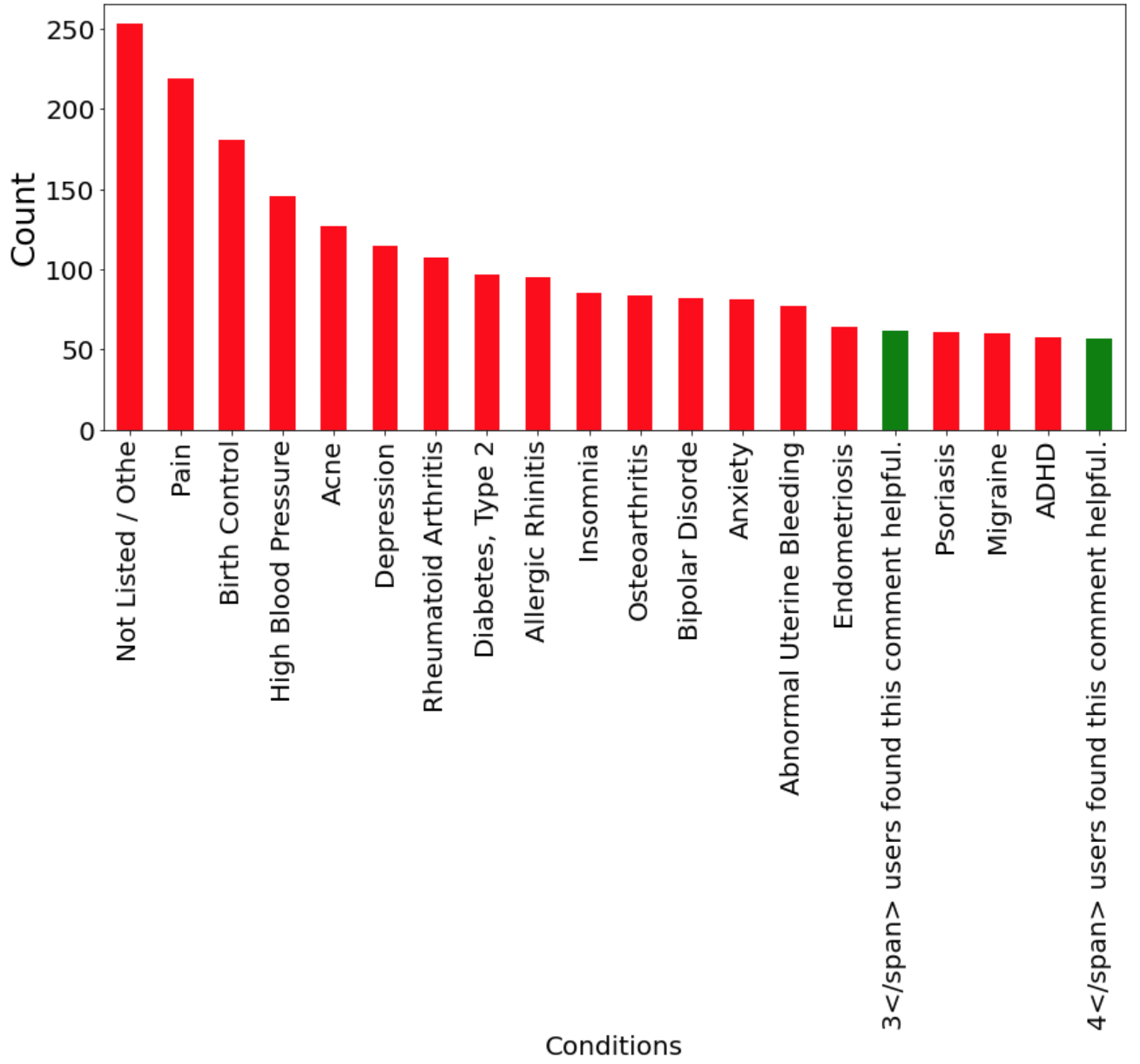}}
\caption{Bar plot of Top 20 conditions that has a maximum number of drugs available}

\end{figure}

Fig. 4 shows the visualization of value counts of the 10-star rating system.  The rating beneath or equivalent to five featured with cyan tone otherwise blue tone. The vast majority pick four qualities; 10, 9, 1, 8, and 10 are more than twice the same number. It shows that the positive level is higher than the negative, and people's responses are polar.
\begin{figure}[htbp]
\centerline{\includegraphics[width=8cm, height=4cm]{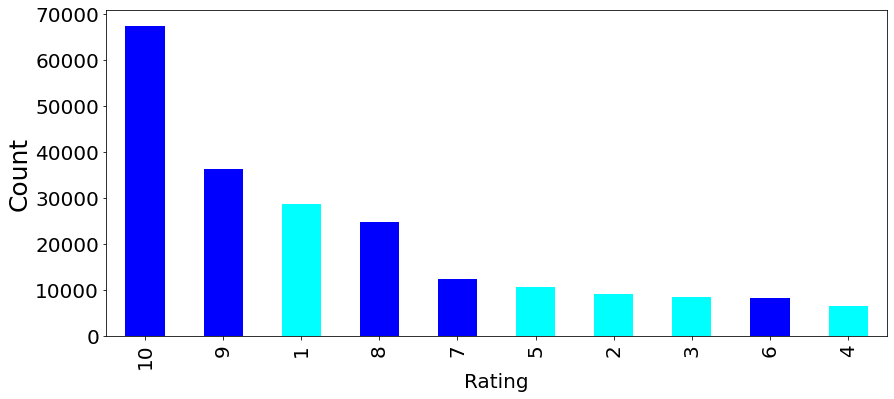}}
\caption{Bar plot of count of rating values versus 10 rating number}

\end{figure}

The condition and drug column were joined with review text because the condition and medication words also have predictive power. Before proceeding to the feature extraction part, it is critical to clean up the review text before vectorization. This process is also known as text preprocessing. We first cleaned the reviews after removing HTML tags, punctuations, quotes, URLs, etc. The cleaned reviews were lowercased to avoid duplication, and tokenization was performed for converting the texts into small pieces called tokens.
Additionally, stopwords, for example, “a, to, all, we, with, etc.,” were removed from the corpus. The tokens were gotten back to their foundations by performing lemmatization on all tokens. For sentiment analysis, labeled every single review as positive and negative based on its user rating. If the user rating range between 6 to 10, then the review is positive else negative.
\subsection{Feature Extraction}
After text preprocessing, a proper set up of the data required to build classifiers for sentiment analysis. Machine learning algorithms can't work with text straightforwardly; it should be changed over into numerical format. In particular, vectors of numbers. A well known and straightforward strategy for feature extraction with text information used in this research is the bag of words (Bow) [16], TF-IDF [17], Word2Vec [18]. Also used some feature engineering techniques to extract features manually from the review column to create another model called manual feature aside from Bow, TF-IDF, and Word2Vec. 
\subsubsection{Bow}
Bag of words [16] is an algorithm used in natural language processing responsible for counting the number of times of all the tokens in review or document. A term or token can be called one word (unigram), or any subjective number of words, n-grams. In this study, (1,2) n-gram range is chosen. Fig. 5 outlines how unigrams, digrams, and trigrams framed from a sentence. The Bow model experience a significant drawback, as it considers all the terms without contemplating how a few terms are exceptionally successive in the corpus, which in turn build a large matrix that is computationally expensive to train.
\begin{figure}[htbp]
\centerline{\includegraphics[width=7cm, height=4cm]{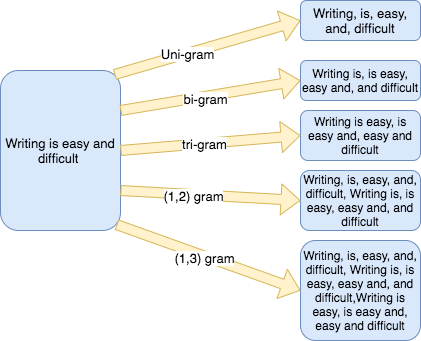}}
\caption{Comparison of various types of grams framed from a sentence}

\end{figure}
\subsubsection{TF-IDF}
TF-IDF [17] is a popular weighting strategy in which words are offered with weight not count. The principle was to give low importance to the terms that often appear in the dataset, which implies TF-IDF estimates relevance, not a recurrence. Term frequency (TF) can be called the likelihood of locating a word in a document. 
\begin{equation}
tf(t,d) = log(1+freq(t,d))
\end{equation}
Inverse document frequency (IDF) is the opposite of the number of times a speciﬁc term showed up in the whole corpus. It catches how a specific term is document speciﬁc. 
\begin{equation}
idf(t,d) = log(\frac{N}{count(d \epsilon D : t \epsilon d)})
\end{equation}
TF-IDF is the multiplication of TF with IDF, suggesting how vital and relevant a word is in the document. 
\begin{equation}
tf idf(t,d,D) = tf(t,d).idf(t,D)
\end{equation}
Like Bow, the selected n-gram range for TF-IDF in this work is (1,2). 
\subsubsection{Word2Vec}
Even though TF and TF-IDF are famous vectorization methods used in different natural language preparing tasks [27], they disregard the semantic and syntactic likenesses between words. For instance, in both TF and TF-IDF extraction methods, the words lovely and delightful are called two unique words in both TF and TF-IDF vectorization techniques although they are almost equivalents.
Word2Vec [18] is a model used to produce word embedding. Word-embeddings reproduced from gigantic corpora utilizing various deep learning models [19]. Word2Vec takes an enormous corpus of text as an input and outputs a vector space, generally composed of hundred dimensions. The fundamental thought was to take the semantic meaning of words and arrange vectors of words in vector space with the ultimate objective that words that share similar sense in the dataset are found close to one another in vectors space.
\subsubsection{Manual Features}
Feature engineering is a popular concept which helps to increase the accuracy of the model. We used fifteen features, which include usefulcount, the condition column which is label encoded using label encoder function from Scikit library, day, month, year features were developed from date column using DateTime function using pandas.
Textblob toolkit [20] was used to extract the cleaned and uncleaned reviews polarity and added as features along with a total of 8 features generated from each of the text reviews as shown in Table I.
\begin{table}[htbp]
\caption{List of Features extracted manually from user reviews}
\begin{center}
\begin{tabular}{|c|c|}
\hline
\textbf{Feature } & \textbf{Description}\\
\hline
Punctuation& Counts the number of punctuation\\
\hline
Word& Counts the number of words\\
\hline
Stopwords& Counts the number of stopwords\\
\hline
Letter& Counts the number of letters\\
\hline
Unique& Counts the number of unique words\\
\hline
Average& Counts the mean length of words\\
\hline
Upper& Counts the uppercase words\\
\hline
Title& Counts the words present in title\\
\hline
\end{tabular}

\end{center}
\end{table}
\subsection{Train Test Split}
We created four datasets using Bow, TF-IDF, Word2Vec, and manual features. These four datasets were split into 75\% of training and 25\% of testing. While splitting the data, we set an equal random state to ensure the same set of random numbers generated for the train test split of all four generated datasets.
\subsection{Smote}
After the Train Test split, only the training data undergone a synthetic minority over-sampling technique (Smote) [22] to prevent the class imbalance problem. Smote is an oversampling technique that synthesized new data from existing data. Smote generates the new minority class data by linear interpolation of randomly selected minority instance 'a' in combination with its k nearest neighbor instance 'b' in the feature space. Table II shows the total distribution of data on final dataset i.e. after data cleaning. Fig. 6 shows the projection of non-smote and smote using t-distributed stochastic neighbor embedding (t-SNE) [21] of 1000 rows on manual features data. It displays that there are more orange points in the non-smote t-SNE projection, which represents the majority class dominance. It also shows that there has been an increment in blue points after using smote that brings out the balance between a majority and minority class that curbs the predominance of the majority class.
\begin{table}[htbp]
\caption{Dataset Distribution}
\begin{center}
\begin{tabular}{|c|c|c|c|}
\hline
\textbf{Smote} & \textbf{Class}& \textbf{Train (75\%)}& \textbf{Test (25\%)}\\
\hline
No& Negative& 47522&15841\\
& Positive& 111583&37195\\
& Total& 159105&53036\\
\hline
Yes& Negative& 78108&15841\\
& Positive& 111583&37195\\
& Total& 189691&53036\\
\hline
\end{tabular}

\end{center}
\end{table}
\begin{figure}[htbp]
\includegraphics[width=4cm, height=4cm]{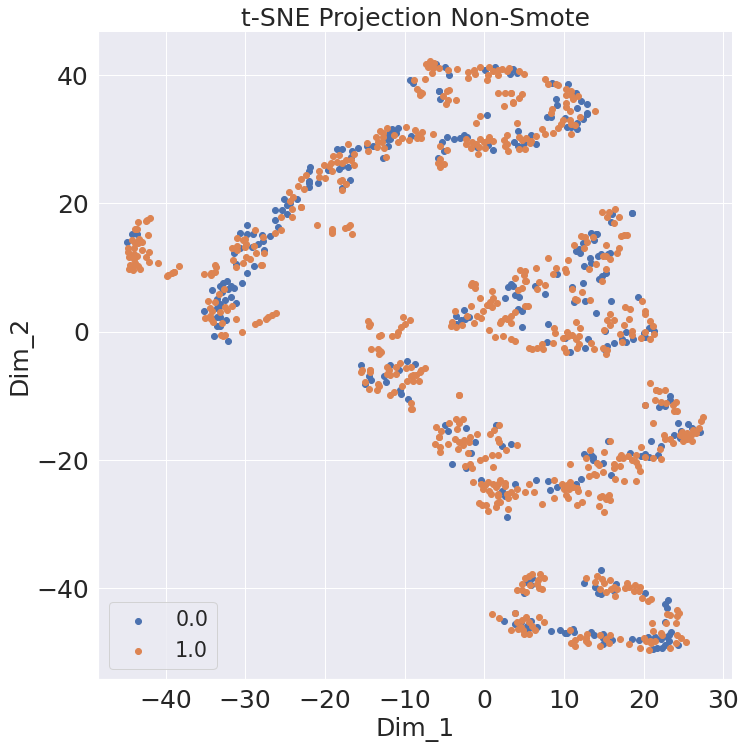}
\includegraphics[width=4cm, height=4cm]{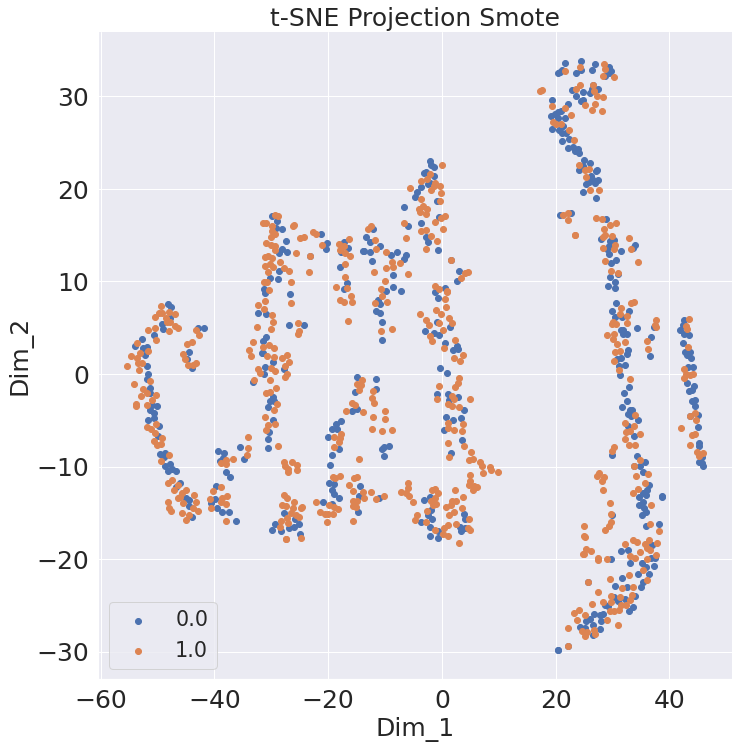}
\caption{t-SNE subplot before and after Smote using 1000 training samples}
\label{fig}
\end{figure}

\subsection{Classifiers}
Distinct machine-learning classification algorithms were used to build a classifier to predict the sentiment. Logistic Regression, Multinomial Naive Bayes, Stochastic gradient descent, Linear support vector classifier, Perceptron, and Ridge classifier experimented with the Bow, TF-IDF model since they are very sparse matrix and applying tree-based classifiers would be very time-consuming. Applied Decision tree, RandomForest, LGBM, and CatBoost classifier on Word2Vec and manual features model. A significant problem with this dataset is around 210K reviews, which takes substantial computational power. We selected those machine learning classification algorithms only that reduces the training time and give faster predictions.
\subsection{Metrics}
The predicted sentiment were measured using five metrics, namely, precision (Prec), recall (Rec), f1score (F1), accuracy (Acc.) and AUC score [23].
Let the letter be: 
$T_p$ = True positive or occurrences where model predicted the positive sentiment truly,
$T_n$ = True negative or occurrences where  model predicted the negative class truly,
$F_p$ = False positive or occurrences where  model predicted the positive class falsely,
$F_n$ = False negative or occurrences where model predicted the negative class falsely,
Precision, recall, accuracy, and f1score shown in equations given below,
\begin{equation}
Precision = \frac{T_{p}}{T_{p}+F_{p}}
\end{equation}
\begin{equation}
\\Recall = \frac{T_{p}}{T_{p}+F_{n}}
\end{equation}
\begin{equation}
Accuracy = \frac{T_{p}+T_{n}}{T_{p}+T_{n} + F_{p}+F_{n}}
\end{equation}
\begin{equation}
F1score = 2.\frac{Precision.Recall}{Precision+Recall}
\end{equation}
Area under curve (Auc) score helps distinguish a classifier's capacity to compare classes and utilized as a review of the region operating curve (roc) curve. Roc curve visualizes the relationship between true positive rate (Tpr) and false positive rate (Fpr) across various thresholds.

\subsection{Drug Recommender system}
After assessing the metrics, all four best-predicted results were picked and joined together to produce the combined prediction. The merged results were then multiplied with normalized useful count to generate an overall score of drug of a particular condition. The higher the score, the better is the drug. The motivation behind the standardization of the useful count was looking at the distribution of useful count in Fig. 7; one may analyze that the contrast among the least and most extreme is around 1300, considerable.
Moreover, the deviation is enormous, which is 36. The purpose behind  is that the more medications individuals search for, the more individuals read the survey regardless of their review is positive or negative, which makes the useful count high. So while building the recommender system, we normalized useful count by conditions.
\begin{figure}[htbp]
\centerline{\includegraphics[width=8cm, height=4cm]{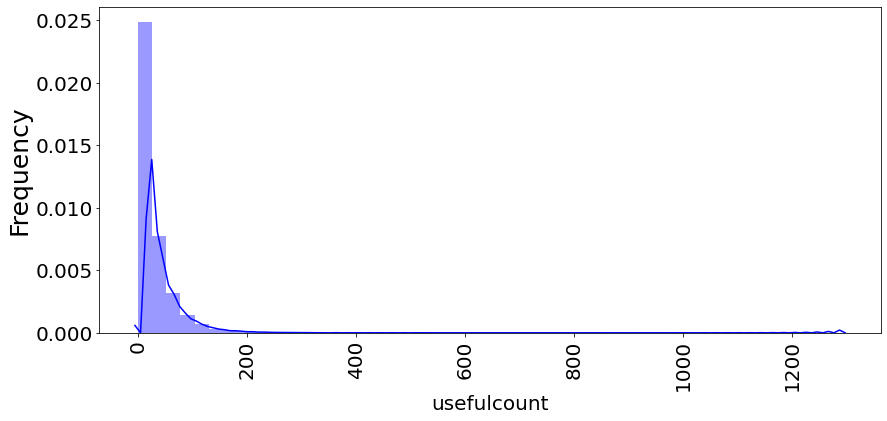}}
\caption{Distribution of Useful Count}

\end{figure}

\section{Results}

In this work, each review was classiﬁed as positive or negative, depending on the user's star rating. Ratings above five are classified as positive, while negative ratings are from one to five-star ratings. Initially, the number of positive ratings and negative ratings in training data were 111583 and 47522, respectively. After applying smote, we increased the minority class to have 70 percent of the majority class examples to curb the imbalances. The updated training data contains 111583 positive classes and 78108 negative classes. Four different text representation methods, namely Bow, TF-IDF, Word2Vec, Manual feature and ten different ML algorithms were applied for binary classification. Results belonging to 5 different metrics given in Tables III, IV, V, VI.

Table  III shows the results using evaluation metrics on a bag of words vectorization technique. We can easily see that perceptron outperforms all other classification algorithms. All algorithms showed similar types of results ranging from 89\% to 91\% accuracy. Logistic regression and LinearSVC accomplished a 90\% AUC score. Even after achieving accuracy more prominent than logistic and LinearSVC,  Perceptron achieved only 89.8\% AUC score.
\begin{table}[htbp]
\caption{bag-of-words}
\begin{center}
\begin{tabular}{|c|c|c|c|c|c|c|}
\hline
\textbf{Model } & \textbf{Class}& \textbf{Prec}& \textbf{Rec}& \textbf{F1}
& \textbf{Acc.} & \textbf{AUC}\\
\hline
LogisticRegression& negative& 0.85&0.87 &0.86&0.91 &0.90 \\
& positive& 0.94&0.93 &0.94&&\\
\hline
Perceptron& negative& 0.87&0.85 &0.86&0.92&0.898\\
& positive& 0.94&0.94 &0.94&& \\
\hline
RidgeClassifier& negative& 0.80&0.87 &0.84&0.90&0.892 \\
& positive& 0.94&0.91 &0.93&&\\
\hline
MultinomialNB& negative& 0.81&0.85 &0.83&0.89&0.881 \\
& positive& 0.93&0.92 &0.92&&\\
\hline
SGDClassifier& negative& 0.80&0.85 &0.82&0.89&0.878 \\
& positive& 0.93&0.91 &0.92&&\\
\hline
LinearSVC& negative& 0.84&0.87 &0.86&0.91&0.90\\
& positive& 0.94&0.93 &0.94&&\\
\hline
\end{tabular}

\end{center}
\end{table}

Table IV manifests the metrics on the TF-IDF vectorization method. LinearSVC increased the TF-IDF vectorization method performance to 93\%, which is more noteworthy than the accuracy achieved by perceptron (91\%) using bag of words model. There was a close competition between LinearSVC, perceptron, and ridge classifier, with only a 1\% difference. However, LinearSVC was picked as the best algorithm since the AUC score is 90.7\%, which is greater than all other algorithms.
\begin{table}[htbp]
\caption{TF-IDF}
\begin{center}
\begin{tabular}{|c|c|c|c|c|c|c|}
\hline
\textbf{Model } & \textbf{Class}& \textbf{Prec}& \textbf{Rec}& \textbf{F1}
& \textbf{Acc.}& \textbf{AUC}\\
\hline
LogisticRegression& negative& 0.79&0.74 &0.76&0.86&0.826 \\
& positive& 0.89&0.92 &0.90&& \\
\hline
Perceptron& negative& 0.89&0.83 &0.86&0.92&0.895\\
& positive& 0.93&0.96 &0.94&& \\
\hline
RidgeClassifier& negative& 0.89&0.84 &0.86&0.92&0.897 \\
& positive& 0.93&0.95 &0.95&&\\
\hline
MultinomialNB& negative& 0.85&0.83 &0.84&0.90&0.883 \\
& positive& 0.93&0.94 &0.93&&\\
\hline
SGDClassifier& negative& 0.76&0.57 &0.65&0.82&0.745 \\
& positive& 0.83&0.92 &0.88&& \\
\hline
LinearSVC& negative& 0.89&0.86 &0.87&0.93&0.907\\
& positive& 0.94&0.96 &0.95&&\\
\hline
\end{tabular}

\end{center}
\end{table}

The performance metrics of various classification methods on Word2Vec can be analyzed using Table V. The best accuracy is 91\% by the LGBM model. Random forest and catboost classifier provide comparable sort of results whereas decision tree classifier performed poorly.
Analyzing the region operating curve score, we can easily manifest that the LGBM has the highest AUC score of 88.3\%.
\begin{table}[htbp]
\caption{Word2Vec}
\begin{center}
\begin{tabular}{|c|c|c|c|c|c|c|}
\hline
\textbf{Model} & \textbf{Class}& \textbf{Prec}& \textbf{Rec}& \textbf{F1}
& \textbf{Acc.}& \textbf{AUC}\\
\hline
DecisionTree& negative& 0.61&0.69 &0.65&0.78&0.751 \\
Classifier& positive& 0.86&0.81 &0.84&&\\
\hline
RandomForest& negative& 0.86&0.77 &0.81&0.89&0.858\\
Classifier& positive& 0.91&0.95 &0.93&& \\
\hline
LGBM& negative& 0.86&0.82 &0.84&0.91&0.883 \\
Classifier& positive& 0.93&0.94 &0.93&&\\
\hline
CatBoost& negative& 0.81&0.79 &0.80&0.88&0.855 \\
Classifier& positive& 0.91&0.92 &0.92&&\\
\hline
\end{tabular}

\end{center}
\end{table}

Table VI displays the performance metrics of four different classification algorithms on manually created features on user reviews. Compared to all other text classification methods, the results are not pretty impressive. However, the random forest achieved a good accuracy score of 88\%.
\begin{table}[htbp]
\caption{Manual Feature Selection}
\begin{center}
\begin{tabular}{|c|c|c|c|c|c|c|}
\hline
\textbf{Model } & \textbf{Class}& \textbf{Prec}& \textbf{Rec}& \textbf{F1}
& \textbf{Acc.}& \textbf{AUC}\\
\hline
DecisionTree& negative& 0.65&0.75 &0.69&0.80&0.816 \\
Classifier& positive& 0.88&0.83 &0.85&& \\
\hline
RandomForest& negative& 0.79&0.81 &0.80&0.88&0.857\\
Classifier& positive& 0.92&0.91 &0.91&& \\
\hline
LGBM& negative& 0.74&0.74 &0.74&0.85&0.787 \\
Classifier& positive& 0.89&0.89 &0.89&&\\
\hline
CatBoost& negative& 0.72&0.73 &0.73&0.84&0.804 \\
Classifier& positive& 0.88&0.88 &0.88&&\\
\hline
\end{tabular}

\end{center}
\end{table}

After evaluating all the models, the prediction results of Perceptron (Bow), LinearSVC (TF-IDF), LGBM (Word2Vec), and RandomForest (Manual Features) was added to give combined model predictions. The main intention is to make sure that the recommended top drugs should be classified correctly by all four models. If one model predicts it wrong, then the drug's overall score will go down. These combined predictions were then multiplied with normalized useful count to get an overall score of each drug. This was done to check that enough people reviewed that drug. The overall score is divided by the total number of drugs per condition to get a mean score, which is the final score. Fig. 8 shows the top four drugs recommended by our model on top five conditions namely, Acne, Birth Control, High Blood Pressure, Pain and Depression.
\begin{figure}[htbp]
\centerline{\includegraphics[width=7cm, height=7cm]{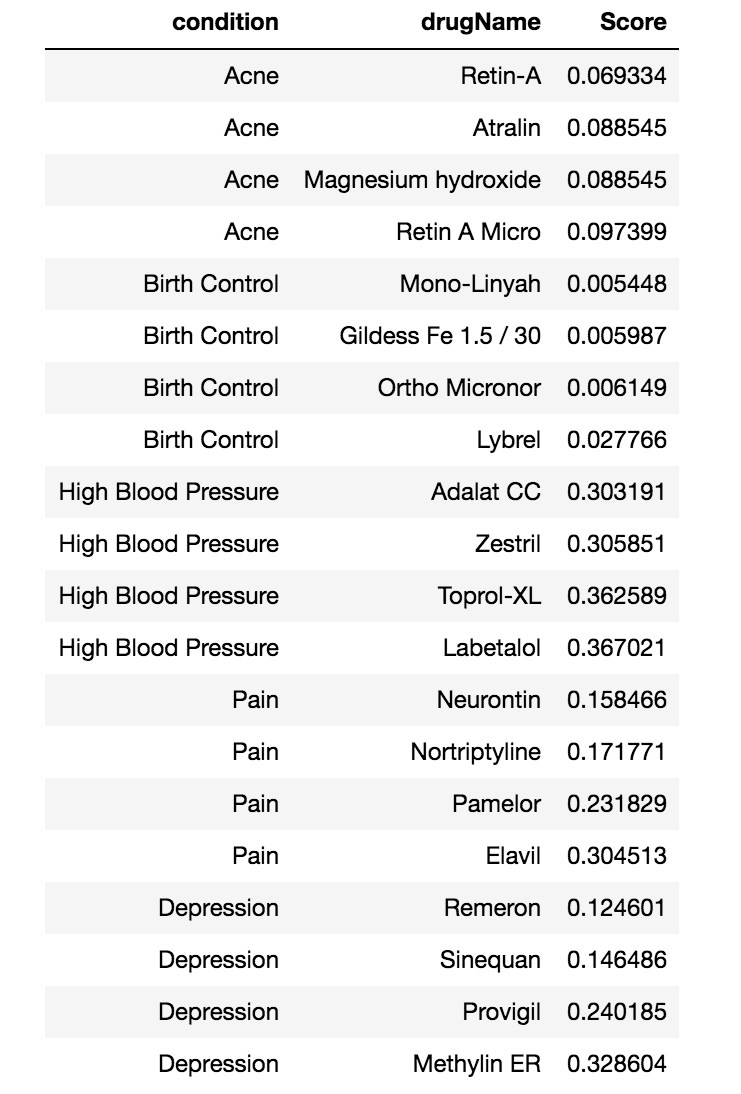}}
\caption{Recommendation of top four drugs on top five conditions}
\end{figure}

\section{Discussion}
The results procured from each of the four methods are good, yet that doesn't show that the recommender framework is ready for real-life applications. It still need improvements. Predicted results show that the difference between the positive and negative class metrics indicates that the training data should be appropriately balanced using algorithms like Smote, Adasyn [24], SmoteTomek [25], etc. Proper hyperparameter optimization is also required for classification algorithms to improve the accuracy of the model.
In the recommendation framework, we simply just added the best-predicted result of each method. For better results and understanding, require a proper ensembling of different predicted results. This paper intends to show only the methodology that one can use to extract sentiment from the data and perform classification to build a recommender system.

\section{Conclusion}
Reviews are becoming an integral part of our daily lives; whether go for shopping, purchase something online or go to some restaurant, we first check the reviews to make the right decisions. Motivated by this, in this research sentiment analysis of drug reviews was studied to build a recommender system using different types of machine learning classiﬁers, such as Logistic Regression, Perceptron, Multinomial Naive Bayes, Ridge classifier, Stochastic gradient descent, LinearSVC, applied on Bow, TF-IDF, and classifiers such as Decision Tree, Random Forest, Lgbm, and Catboost were applied on Word2Vec and Manual features method. We evaluated them using five different metrics, precision, recall, f1score, accuracy, and AUC score, which reveal that the Linear SVC on TF-IDF outperforms all other models with 93\% accuracy. On the other hand, the Decision tree classifier on Word2Vec showed the worst performance by achieving only 78\% accuracy. We added best-predicted emotion values from each method, Perceptron on Bow (91\%), LinearSVC on TF-IDF (93\%), LGBM on Word2Vec (91\%), Random Forest on manual features (88\%), and multiply them by the normalized usefulCount to get the overall score of the drug by condition to build a recommender system. Future work involves comparison of different oversampling techniques, using different values of n-grams, and optimization of algorithms to improve the performance of the recommender system.

\end{document}